\documentclass[a4paper,10pt]{article}

\usepackage[dvips]{graphicx}

\usepackage{amssymb}
\usepackage{amsmath}
\usepackage{latexsym}
\usepackage{array}
\usepackage{graphicx}

\usepackage{theorem}
\theoremstyle{break}
\newtheorem{Theorem}{Theorem}
\newtheorem{Proposition}{Proposition}
\newtheorem{Definition}{Definition}
\newtheorem{Corollary}{Corollary}
\newtheorem{Example}{Example}
\newtheorem{Lemma}{Lemma}
\newtheorem{Conjecture}{Conjecture}
\def\qed{\hfill\hbox{$\Box$}\vspace{10pt}\break}

\begin{document}
\title{Graphs emerging from the solutions to the periodic discrete Toda equation over finite fields}

\author{Masataka Kanki${}^1$, Y\={u}ki Takahashi${}^2$, and Tetsuji Tokihiro${}^3$ \\
\small
${}^1$ Faculty of Engineering Science,\\
\small Kansai University, 3-3-35 Yamate, Suita, Osaka 564-8680, Japan\\
\small ${}^2$ Department of Mathematics,\\
\small University of California, Irvine, Irvine CA 92697, USA\\
\small ${}^3$ Graduate School of Mathematical Sciences,\\
\small University of Tokyo, 3-8-1 Komaba, Tokyo 153-8914, Japan
}

\date{}

\maketitle

\begin{abstract} 
The periodic discrete Toda equation defined over finite fields has been studied.
We obtained the finite graph structures constructed by the network of states where edges denote possible time evolutions.
We simplify the graphs by introducing a equivalence class of cyclic permutations to the initial
values. We proved that the graphs are bi-directional and that they are composed of several arrays of complete graphs connected at one of their vertices.
The condition for the graphs to be bi-directional is studied for general discrete equations.

\end{abstract}

\maketitle

\section{Introduction}
The Toda lattice has first been introduced as a mechanical model of the chain of particles with nonlinear interaction force by springs \cite{TODA}. Toda has discovered that the nonlinear system
\[
m \frac{d^2 u_n}{dt^2}=\phi '(r_{n+1})-\phi'(r_n),
\]
with an exponential potential (which is now called the Toda potential)
\[
\phi(r)=\frac{a}{b}e^{-br}+ar,
\]
is exactly solvable. Here, $u_n$ is the position of the $n$th particle, $r_n=u_n-u_{n-1}$ is the stretch of the spring connecting the particles, and $a,b$ are arbitrary parameters with $ab>0$.
It is an important example of nonlinear systems which has analytic solutions just like linear systems.
Hirota and Suzuki have constructed an electrical circuit which simulates the behavior of the Toda lattice \cite{HS}.
Time discretization (fully-discretized version) of the Toda lattice has been obtained \cite{H,H2}.  Its generalization and the proof of complete integrability has been done by Suris \cite{suris,suris2}. We focus on the fully-discretized Toda lattice equation in \cite{H2} in this article, and study the behavior of the solutions of the equation over finite fields.
Study of the integrable equations over finite fields is of importance for the following reasons. First, as each variable is allowed to take only finite number of values, it is easy for us to obtain the result numerically without errors. Second, the system over finite fields can be naturally seen as an analogue of a {\em cellular automaton}. The cellular automaton is a discrete dynamical system the cell of which takes only a finite number of states \cite{W}. Studying the discrete systems over finite fields can provide fundamental models in analyzing cellular automata and so-called {\em ultra-discrete} systems.
However, the time evolution of the system is not always well-defined over
finite fields: we observe that the evolution comes to a stop as division by $0 \mod p$ or indeterminacies appear for most of the initial conditions.
To study the system over finite fields we have either to rewrite the equations so that they do not have division terms, or to extend the domain of definition so that the equation is well-defined.
The latter approach of extending the domains is studied in connection with the dynamics over the field of $p$-adic numbers in one-dimensional case. In particular, discrete versions of the Painlev\'{e} equations has been studied \cite{KMTT}.
Na\"{i}ve application of this method to two-dimensional lattices such as the discrete Toda equation, discrete KdV equation has a certain difficulty,
because they have infinitely many singular patterns.
In this article we adopt the former approach and aim to understand the structure of the solutions of two-dimensional lattice equations over finite fields.
In the following sections, we consider the time and space discrete Toda lattice equation over the finite field $\mathbb{F}_p:=\{0,1,2,\cdots,p-1\}$, where $p\ge 2$ is a prime number. 
The discrete Toda equation over finite fields has been investigated by one of the authors, and there the solutions for several symmetric initial conditions have been obtained \cite{T}. This article further generalizes his result and presents the graph structures of the solutions for general initial conditions. The states of the discrete Toda equations are connected by the time evolution, and the pairs of states (vertices) and the connections (edges) form finite graph structures.
One of our results is that this graph is always bi-directional, if we consider the equivalence class of states in terms of cyclic permutations of the variables. The other result is that this graph is composed of finite number of complete graphs (polygons with all the vertices) that are connected by one of their edges.
We have presented several orbits for small primes and small sizes of the system, and have also given a conjecture on the connected components of the graph for $N=2$.
In the last section, we study in a generalized settings, which include the discrete Toda equation, and obtain a sufficient condition for the graphs to be bi-directional.

\section{Time evolution of discrete Toda equation}
In this paper we use the following coupled form of the discretized Toda lattice equation for variables $I_n^t$ and $V_n^t$:
\begin{equation}
\left\{
\begin{array}{cl}
I_n^t+V_n^t&=I_n^{t+1}+V_{n-1}^{t+1},\vspace{2mm} \\
I_{n+1}^t V_n^t&=I_n^{t+1} V_n^{t+1},
\end{array}
\right.
\label{toda}
\end{equation}
where the independent variables $t$ and $n$ take only integer values \cite{H2}.
We deal with Eq. \eqref{toda} over the finite field $\mathbb{F}_p$, where $p$ is a prime number. We omit `mod $p$' for simplicity.
We take $N\ge 1$ as the size of the system and impose the periodic boundary conditions
\begin{equation}
I_{n+N}^t=I_n^t,\ \  V_{n+N}^t=V_n^t, \label{periodic}
\end{equation}
for all $n\in \mathbb{Z}$.
Equation \eqref{toda} has the matrix representation (the so-called Lax representation):
\[
L^{t+1} R^{t+1} =R^t L^t,
\]
where $R^t$ and $L^t$ are the following $N\times N$ matrices:
\[
L^t=
\begin{pmatrix}
1 & 0 & & & \frac{V_N^t}{y}\\
V_1^t & 1 & 0 & & \\
 & V_2^t & 1 & \ddots & \\
 & &  \ddots & \ddots & 0\\
 & &             &  V_{N-1}^t & 1 
\end{pmatrix}
, R^t=
\begin{pmatrix}
I_1^t & 1 & & & \\
 & I_2^t & 1 & & \\
 &  & \ddots & \ddots & \\
 & &   & I_{N-1}^t & 1\\
y & &             &   & I_N^t 
\end{pmatrix}
.
\]
Here, $y$ is an independent variable, which is called the {\em spectral parameter}.
Let us define $f(x,y)=\det (xE+R^t L^t)$. Polynomial $f(x,y)$ is invariant under the shift of $t$, since we have
\begin{eqnarray*}
&&\det (xE+R^t L^t)=\det (xE +L^{t+1} R^{t+1}) \\
&=&\det\left( (L^{t+1})^{-1} (xE+L^{t+1} R^{t+1})L^{t+1}\right)=\det  (xE+R^{t+1} L^{t+1}),
\end{eqnarray*}
for arbitrary value $x$.
The function $f(x,y)$ is called the {\em spectral curve} of the discrete Toda equation \eqref{toda}.
In particular, if we take $x=0$, $\det (R^t L^t)=\det R^t \cdot \det L^t$ is conserved under the time evolution.
From
\[
\det R^t \cdot \det L^t=(-1)^{N-1}y+\frac{(-1)^{N-1}}{y}\prod_{i=1}^{N} (I_i^t V_i^t)+\prod_{i=1}^N I_i^t +\prod_{i=1}^N V_i^t,
\]
we immediately conclude that
\begin{equation}
c_1=\prod_{i=1}^N I_i^t +\prod_{i=1}^N V_i^t,\;\;\mbox{and}\;\; c_2=\prod_{i=1}^{N} (I_i^t V_i^t), \label{cons}
\end{equation}
are the conserved quantities.
The {\em state} of Eq. \eqref{toda} is expressed as a $2N$-dimensional vector:
\[
(I_1^t,V_1^t,I_2^t,V_2^t,\cdots, I_N^t, V_N^t) \in \mathbb{F}_p^{2N}.
\]
In some parts of this article we limit ourselves to $(I_1^t,V_1^t,I_2^t,V_2^t,\cdots,I_N^t, V_N^t)$ $\in \left(\mathbb{F}_p^{\times}\right)^{2N}$, which will be called {\em non-zero states}.
The state is often abbreviated as $(I_1^t,V_1^t,I_2^t,V_2^t, \cdots, I_N^t, V_N^t)=(\mathbb{I}^t\mathbb{V}^t)$.
Note that if we have $(\mathbb{I}^t \mathbb{V}^t)\in \left(\mathbb{F}_p^{\times}\right)^{2N}$ for some $t$, it is clear that $(\mathbb{I}^s\mathbb{V}^s)\in \left(\mathbb{F}_p^{\times}\right)^{2N}$ for all $s>t$.
Therefore the orbits of non-zero states and those of states with at least one zero in their components do not intersect with each other.
\begin{Definition}
We say that the state $(\mathbb{I}^{t+1}\mathbb{V}^{t+1})$ is a next step of $(\mathbb{I}^{t}\mathbb{V}^{t})$,
and use the arrow $\rightarrow$ as
\[
(\mathbb{I}^t\mathbb{V}^t)\rightarrow (\mathbb{I}^{t+1}\mathbb{V}^{t+1}),
\]
when $(\mathbb{I}^t\mathbb{V}^t)$ and $(\mathbb{I}^{t+1}\mathbb{V}^{t+1})$ satisfy Eq. \eqref{toda}.
\end{Definition}

Note that we can have multiple next steps for one particular state.
\begin{Example}
As an example, let us classify the orbits in the case of $p=2$ and $N=2$.
Initial states are $16$ points $(I_1^0,V_1^0,I_2^0,V_2^0)\in(\mathbb{F}_2)^4=\{0,1\}^4$.
We have $(0,0,0,0)\rightarrow (0,0,0,0)$, $(0,0,0,0)\rightarrow (1,0,0,1)$, $(0,0,0,0)\rightarrow (0,1,1,0)$,
$(0,0,0,1)\rightarrow (0,0,1,0)$, $(0,0,0,1)\rightarrow (0,1,0,0)$, $\cdots$.
By investigating all the next steps for $16$ points, we obtain the following orbits in Fig. \ref{fig00}.
We can observe that even for the simplest case, the graphs exhibit complex structure with multiple evolutions and self-loops.
\begin{figure}
\centering
\includegraphics[width=9cm,bb=100 350 350 750]{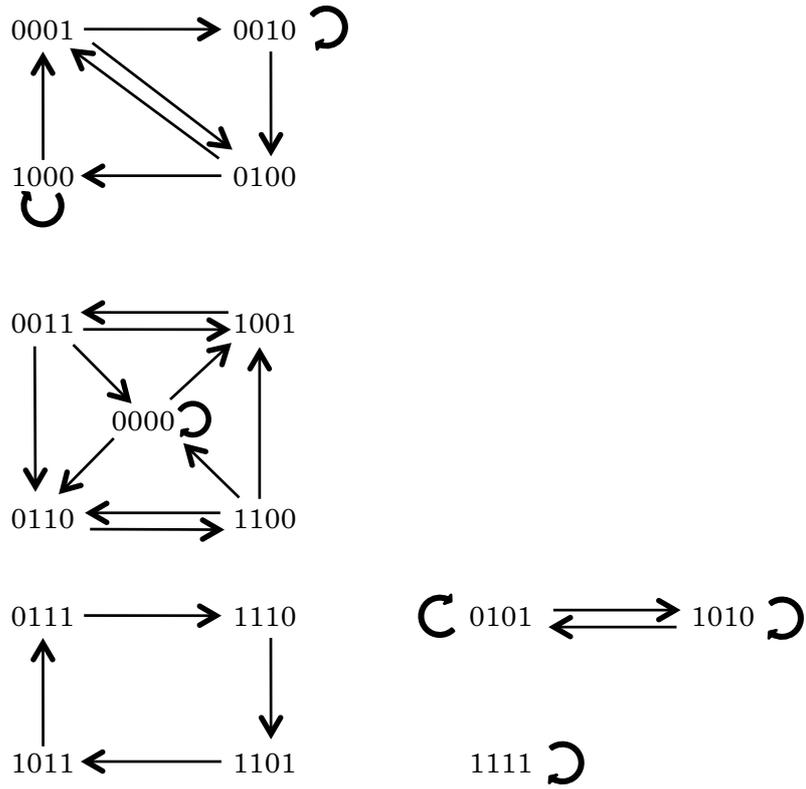}
\caption{Orbits of solutions for all the $16$ states in the case of $p=2,N=2$.
Here the point $\{I_1,V_1,I_2,V_2\}$ is denoted by $I_1 V_1 I_2 V_2$ and is placed at the vertex of the graph.}
\label{fig00}
\end{figure}
\end{Example}
From here on in this section, let us limit ourselves to the case of non-zero states
\[
(I_1^0,V_1^0,I_2^0,V_2^0,\cdots, I_N^0, V_N^0) \in (\mathbb{F}_p^{\times})^{2N}
\]
and study the number of possible next steps.
We define $X_n^t$ and $\alpha^t$ as
\[
X_n^t=1+\frac{V_{n-1}^t}{I_{n-1}^t}+\frac{V_{n-1}^t V_{n-2}^t}{I_{n-1}^t I_{n-2}^t}+\cdots +\frac{\prod_{i=1}^{N-1} V_{n-i}^t}{\prod_{i=1}^{N-1} I_{n-i}^t},
\]
\[
\alpha^t=\frac{V_n^t V_{n-1}^t \cdots V_{n+1}^t}{I_n^t I_{n-1}^t \cdots I_{n+1}^t}=\frac{\prod_{i=1}^N V_i^t}{\prod_{i=1}^N I_i^t},
\]
where the subscripts are considered modulo $N$. From the periodicity, we have $V_{n-(N-1)}^t=V_{n+1}^t$ and $I_{n-(N-1)}^t=I_{n+1}^t$.
Note that, despite the conserved quantities in \eqref{cons}, $\alpha^t$ depends on the time $t$.
In fact, $\alpha^t$ is one of the two solutions of
\begin{equation}
\Lambda^2-c_1 \Lambda +c_2=0. \label{alphaeq}
\end{equation}
The following relation is important:
\begin{equation}
X_{n+1}^t-1+\alpha^t=\frac{V_n^t}{I_n^t}X_n^t. \label{relationofx}
\end{equation}
If $\alpha^t = 1$, we have from \eqref{relationofx} that
\begin{eqnarray}
X_1^t= 0 \ &\Longleftrightarrow& X_n^t= 0 \ \mbox{for all}\  n\in\{1,2,\cdots, N\}, \label{X1}\\
X_1^t\not= 0\ &\Longleftrightarrow& X_n^t\not = 0 \ \mbox{for all}\  n\in\{1,2,\cdots, N\}. \label{X1not}
\end{eqnarray}
Having these relations in mind, we obtain the following proposition.
\begin{Proposition}\label{p-1}
For each $(\mathbb{I}^t\mathbb{V}^t)\in\left(\mathbb{F}_p^{\times}\right)^{2N}$,
we define $\nu_{( I^t V^t )}$ as the number of $(\mathbb{I}^{t+1}\mathbb{V}^{t+1})$'s which are next steps of $(\mathbb{I}^t\mathbb{V}^t)$.
Then we have:
\begin{itemize}
\item For $\alpha^t= 1$, we always have $\alpha^{t+1}=1$, and:
\begin{itemize}
\item If $X_1^t = 0$, then $1\le \nu_{(I^tV^t)} \le p-1$,
\item If $X_1^t \not = 0$, then $\nu_{(I^tV^t)} = 1$.
\end{itemize}
\item For $\alpha^t\not = 1$ we have the following two cases:
\begin{itemize}
\item If $X_n^t \not = 0$ for all $n$, then $\nu_{({I}^t{V}^t)} = 2$, and $\alpha^{t+1}\in\{\alpha^t, (\alpha^t)^{-1} \}$,
\item Otherwise $\nu_{({I}^t{V}^t)} = 1$ and $\alpha^{t+1}=(\alpha^t)^{-1}$.
\end{itemize}
\end{itemize}
\end{Proposition}
\textbf{proof}\;\;
Let us abbreviate $\nu_{({I}^t{V}^t)}$ as $\nu$ in the proof.
Let us first prove that $\nu\le p-1$ for all cases.
Fix $V_1^{t+1}:=j\ \left(j\in\mathbb{F}_p^{\times}\right)$ and check whether other variables are well-defined.
We obtain
\begin{equation}
\left(I_n^t+V_n^t-V_{n-1}^{t+1}\right)\cdot V_n^{t+1}=I_{n+1}^t V_n^t\ \ (n=1,2,\cdots, N) \label{j}
\end{equation}
from equation \eqref{toda}.
By substituting $n=2$ in \eqref{j} we have
\[
V_2^{t+1}=\frac{I_3^t V_2^t}{I_2^t+V_2^t-j}\in \mathbb{F}_q^{\times},
\]
on condition that $I_2^t+V_2^t-j\not = 0$.
If $I_2^t+V_2^t-j = 0$, on the other hand, we do not have $(\mathbb{I}^{t+1} \mathbb{V}^{t+1})$ with $V_1^{t+1}=j$,
since \eqref{j} becomes $0=I_{n+1}^t V_n^t\neq 0$, which is a contradiction.
We conclude by induction that there exists at most one $(\mathbb{I}^{t+1} \mathbb{V}^{t+1})$ that satisfies \eqref{toda} for each $j$.
Therefore $\nu\le \#\mathbb{F}_p^{\times}=p-1$ is proved. Next, we have from \eqref{toda} that
\begin{eqnarray*}
I_n^{t+1}&=&I_n^t+V_n^t-V_{n-1}^{t+1}=I_n^t+V_n^t-\frac{I_n^t V_{n-1}^t}{I_{n-1}^{t+1}}\\
&=&I_n^t+V_n^t-\frac{I_n^t V_{n-1}^t}{I_{n-1}^{t}+V_{n-1}^t-\frac{I_{n-1}^t V_{n-2}^t}{I_{n-2}^{t+1}}}\\
&=&\cdots\\
&=&I_n^t+V_n^t-\frac{I_n^t V_{n-1}^t}{I_{n-1}^{t}+V_{n-1}^t-\frac{I_{n-1}^t V_{n-2}^t}{I_{n-2}^{t}+V_{n-2}^t-\frac{\ddots}{I_{n-N}^{t+1}}}},
\end{eqnarray*}
where we have continued the iteration until we meet $I_{n-N}^{t+1}$.
Since $I_{n-N}^{t+1}=I_n^{t+1}$ from the periodic boundary condition \eqref{periodic},
we can solve the equation above in terms of $I_n^{t+1}$ to obtain
\begin{equation}
X_n^t\left(I_n^{t+1}\right)^2-\left(I_n^t X_{n+1}^t+ V_n^t X_n^t\right) I_n^{t+1}+I_n^tV_n^tX_{n+1}^t=0.
\end{equation}
This quadratic equation can be factorized as
\begin{equation}
\left(I_n^{t+1}-V_n^t\right)\left(X_n^t I_n^{t+1}- X_{n+1}^t I_n^t\right)=0. \label{niji}
\end{equation}
If $\alpha^t= 1$ then we have from \eqref{X1} and \eqref{X1not} that either
(i) ($X_n^t= 0$ for all $n$) or (ii) ($X_n^t\not = 0$ for all $n$).
In the case (i), Eq.~\eqref{niji} is trivially satisfied.
In this case, we can only state that $\nu\le p-1$, since there are no other restrictions.
In the latter case (ii), we have only one solution (double root) for \eqref{niji}. This is because we have, as the second solution of \eqref{niji}, $I_n^{t+1}=\frac{X_{n+1}^t}{X_n^t}I_n^t=V_n^t$, since $X_{n+1}^t I_n^t = X_n^t V_n^t$ from \eqref{relationofx}.
In both cases, since $\alpha^t=1$, Eq. \eqref{alphaeq} has a double root so that $\prod_{i=1}^N V_i^{t+1}=\prod_{i=1}^N I_i^{t+1}=\prod_{i=1}^N V_i^{t}=\prod_{i=1}^N I_i^{t}$, which proves that $\alpha^{t+1}=1$.

Next we study the case where $\alpha^t \not = 1$.
From $c_1=(\alpha^t+1)\prod_{i=1}^N I_i^t$ and $c_2=\alpha^t \left( \prod_{i=1}^N I_i^t \right)^2$,
the solutions of \eqref{alphaeq} are
\[
\Lambda=\alpha^t \prod_{i=1}^N I_i^t\left(=\prod_{i=1}^N V_i^t \right),\;\; \Lambda=\prod_{i=1}^N I_i^t.
\]
Thus
\[
\left( \prod_{i=1}^N I_i^{t+1}, \prod_{i=1}^N V_i^{t+1} \right)=\left( \prod_{i=1}^N I_i^{t}, \prod_{i=1}^N V_i^{t} \right),\ \left( \prod_{i=1}^N V_i^{t}, \prod_{i=1}^N I_i^{t} \right),
\]
which indicates that $\alpha^{t+1}=\alpha^t$ or $\alpha^{t+1}=1/\alpha^t$.
From the relation \eqref{relationofx}, at least one of $X_n^t$ or $X_{n+1}^t$ must be non-zero.
If $X_n^t\not = 0$ and $X_{n+1}^t\not = 0$, we have $X_{n+1}^t I_n^t \not = X_n^t V_n^t$, again by using \eqref{relationofx}. Thus, Eq. \eqref{niji} does not have a double root, and the number of next steps is $\nu= 2$.
If ($X_n^t\equiv 0 $ and $X_{n+1}^t\not\equiv 0$) or ($X_n^t\not\equiv 0 $ and $X_{n+1}^t\equiv 0$) then \eqref{niji} has only one solution: $I_n^{t+1}=V_n^t$.
\qed
\begin{Corollary}
If $\alpha^0$ is well-defined and $\alpha^0\not =0$, then the set
\[
A:=\left\{\alpha^t ,\frac{1}{\alpha^t}\right\}
\]
is conserved under $t$.
\end{Corollary}
Therefore the solutions of \eqref{toda} over $(\mathbb{F}_p^{\times})^{2N}$ are decomposed into at least $(p-1)/2$ orbits, depending on $A$.
We have learned that the special solutions of multiple arrows, which are typical over finite fields,
appear when
\begin{equation} \label{typicalff}
A = \{1\}\ \mbox{and}\ (\forall n, \ X_n^t = 0\ ).
\end{equation}
 
\begin{Example}\label{rei1}
Let $p=7$ and $N=2$.
The state $(I_1^{t+1},V_1^{t+1},I_2^{t+1},V_2^{t+1})=(1,3,5,1)$ is a next step of $(I_1^{t},V_1^{t},I_2^{t},V_2^{t})=(3,6,4,4)$: i.e.,
\[
(3,6,4,4) \rightarrow (1,3,5,1).
\]
We also have
\[
(3,6,4,4) \rightarrow (6,4,4,3).
\]
In this case $\alpha^t\equiv 2\!\!\mod 7$ and
\[
X_1^t=1+\frac{V_1^t}{I_1^t}=1+2=3\not \equiv 0,\ \ X_2^t=1+1=2\not\equiv 0 \mod 7
\]
and $\nu_{(3,6,4,4)}=2$.
We have $\alpha^{t+1}\equiv 2\!\!\mod 7$ for $(1,3,5,1)$, and $\alpha^{t+1}\equiv 12/24\equiv 1/2\equiv 4\!\!\mod 7$ for $(6,4,4,3)$. The set $A=\{2,4\}$ is independent of $t$.
\end{Example}
\begin{Example} \label{ex133343}
Let $p=5$ and $N=3$.
We consider the state $(I_1^t,V_1^t,I_2^t,V_2^t,I_3^t,V_3^t)=(1,3,3,3,4,3)$.
In this case $\alpha^t \equiv 1\mod 5$ and
\[
X_1^t=1+\frac{V_3^t}{I_3^t}+\frac{V_3^t V_2^t}{I_3^t I_2^t}=1+\frac{3}{4}+\frac{9}{12}\equiv 1+2+2\equiv 0\mod 5.
\]
We also have $X_2^t\equiv X_3^t\equiv 0\mod 5$.
The next steps of $(1,3,3,3,4,3)$ are the following three states:
\[
(3,3,3,4,3,1),\ (2,2,4,3,4,2),\ (1,4,2,1,1,3).
\]
\end{Example}
Example \ref{ex133343} is one of the cases of condition \eqref{typicalff}.


\section{Equivalence class of initial values}\label{sec3}
In this section, we do not limit ourselves to non-zero initial conditions.
The equation \eqref{toda} is always satisfied when
\begin{equation}
\left\{
\begin{array}{cl}
I_n^{t+1}&=V_n^t,\vspace{2mm} \\
V_n^{t+1}&=I_{n+1}^t.
\end{array}
\right. (n=1,2,\cdots, N) \label{trivial}
\end{equation}
\begin{Definition}
We call the above evolution \eqref{trivial} from $(\mathbb{I}^t\mathbb{V}^t)$ to $(\mathbb{I}^{t+1}\mathbb{V}^{t+1})$ the {\em trivial evolution} and denote it by
\[
\text{triv} (\mathbb{I}^t\mathbb{V}^t)=(\mathbb{I}^{t+1}\mathbb{V}^{t+1}).
\]
\end{Definition}
For example, the state $(6,4,4,3)$ in example \ref{rei1} is a trivial evolution of $(3,6,4,4)$: i.e.,
triv$(3,6,4,4)=(6,4,4,3)$.
\begin{Definition}\label{equiv}
We define a equivalence relation $\sim$ as follows:
\[
(\mathbb{I}\mathbb{V})\sim (\mathbb{I'}\mathbb{V'}) \Longleftrightarrow \exists\  i\in\mathbb{Z} \text \ {s.t.}\  \left(\text{triv}\right)^i (\mathbb{I}\mathbb{V})=(\mathbb{I'}\mathbb{V'})
\]
\end{Definition}
Note that (triv)$^{2N}$ is an identity map over $\mathbb{F}_p^{2N}$.
We define the set of equivalence classes as
\[
\Omega(p,N) :=\left(\mathbb{F}_p\right)^{2N}/\sim.
\]
We will well-define the network of evolutions of the periodic discrete Toda equation \eqref{toda} over $\Omega(p,N)$ in the next section.
Note that we have the following bijective mapping $\phi$ between $\left(\mathbb{F}_p\right)^{2N}$
and a set of integers $S:=\{0,1,\cdots ,p^{2N}-1\}$:
\begin{equation}
\begin{array}{ccc}
\phi: \left(\mathbb{F}_p\right)^{2N} &\longrightarrow &S\vspace{2mm} \\
(I_1,V_1,\cdots, I_N,V_N) &\mapsto &I_1 p^{2N-1}+V_1 p^{2N-2}+\cdots+I_N p+V_N.
\end{array}
\end{equation}
In short we are interpreting the element of $\left(\mathbb{F}_p\right)^{2N}$ as a $p$-adic number.
Therefore, Eq. \eqref{toda} is well-defined as a relation in $S$.
The equivalence relation over $S$ is naturally induced from that over $\left(\mathbb{F}_p\right)^{2N}$.
Therefore we can identify $(S/ \sim)$ with $\Omega$.
By utilizing the set $S$, we can simplify the expression of the evolution and facilitate numerical experiments.
\begin{Example}
Let $p=7$ and $N=2$. Since $\phi((3,6,4,4))=1355\in S$, $\phi((1,3,5,1))=526$ and $\phi((6,4,4,3))=2285$,
the first evolution in example \ref{rei1} is expressed as $1355\rightarrow 526$ and the second one as $1355\rightarrow 2285$ as elements of $S$.
As an element of $\Omega(7,2)$, we have
\[
\Omega(7,2)\ni \left[(3,6,4,4)\right]=\left\{(3,6,4,4),(6,4,4,3),(4,4,3,6),(4,3,6,4) \right\},
\]
and this equivalence class corresponds to the following class in $S/\sim$:
\[
S/\sim\ \ni \left[\phi((3,6,4,4))\right]=[1355]=(1355,2285,1595,1565).
\]
\end{Example}


\section{Graph structures}\label{sec4}
Let us use $\Omega$ instead of $\Omega(p,N)$ in this section, since the discussion does not depend on $p$ and $N$.
We identify $\Omega$ with $S/ \sim$.
Therefore $(\mathbb{IV})=(I_1,V_1,\cdots, I_N,V_N)\in (\mathbb{F}_p)^{2N}$ is identified with a scalar
$\phi((\mathbb{IV}))=I_1 p^{2N-1}+\cdots + V_N\in S$.
The trivial evolution (triv) is naturally induced over $S$: i.e., for $x\in S$, triv$(x)$ is defined as $\phi\left(\mbox{triv} \left(\phi^{-1}(x)\right)\right)\in S$.
\begin{Definition}
For two equivalence classes $A,\ B\in\Omega$, we say that we have an {\em evolution} from $A$ to $B$ when
there exist $a\in A$ and $b\in B$ such that $a\rightarrow b$.
\end{Definition}
When we have an evolution from $A$ to $B$, we denote the evolution as $A\Rightarrow B$.
\begin{Proposition}\label{abc}
For $A,B,C \in \Omega$, we assume that there exist $a\in A,\ b\in B$ and $c\in C$ such that we
have $a\rightarrow b$ and $a\rightarrow c$.
Then we have $B\Rightarrow C$.
\end{Proposition}
\textbf{proof}

We construct an evolution from an element of $B$ to that of $C$.
Let us assume that $a=(\mathbb{I} \mathbb{V})$, $b=(\mathbb{I}' \mathbb{V}')$ and $c=(\mathbb{I}'' \mathbb{V}'')$.
We consider
\[
(\text{triv})^{-1}(\mathbb{I}' \mathbb{V}')=(V'_N,I'_1,V'_1,I'_2,\cdots ,V'_{N-1},I'_N),
\]
and rename this point as
\[
(I'''_1,V'''_1,I'''_2,V'''_2,\cdots ,V'''_N,I'''_N)=(V'_N,I'_1,V'_1,I'_2,\cdots ,V'_{N-1},I'_N).
\]
Then
\begin{equation}
\left\{
\begin{array}{cl}
I'''_i+V'''_{i}=V'_{i-1}+I'_{i},\vspace{2mm} \\
I'''_{i+1}V'''_i=V'_{i}I'_{i}.
\end{array}
\right. (i=1,\cdots,N) \label{trivinverse}
\end{equation}
Since $a\rightarrow b$ and $a\rightarrow c$, we have
\begin{equation}
\left\{
\begin{array}{ccl}
I_i+V_{i}&=I'_{i}+V'_{i-1}&=I''_{i}+V''_{i-1},\vspace{2mm} \\
I_{i+1}V_i&=I'_{i}V'_i&=I''_{i}V''_{i}.
\end{array}
\right. (i=1,\cdots, N) \label{atobc}
\end{equation}
From \eqref{trivinverse} and \eqref{atobc} we obtain
\begin{equation}
\left\{
\begin{array}{cl}
I'''_i+V'''_{i}=I''_{i}+V''_{i-1},\vspace{2mm} \\
I'''_{i+1}V'''_i=I''_{i}V''_{i},
\end{array}
\right. (i=1,\cdots,N)
\end{equation}
which indicates that
\[
(\text{triv})^{-1}\left(b\right)\rightarrow c.
\]
Therefore $B\Rightarrow C$.
\qed

Note that $A\Rightarrow B$ and $A\Rightarrow C$ are not sufficient to prove that $B\Rightarrow C$.
We could state that proposition \ref{abc} is a weak type transitive law.

\begin{Proposition}
For $A,B\in \Omega$, if $A\Rightarrow B$ then $B\Rightarrow A$. \label{bidirection}
\end{Proposition}
\textbf{proof}

Since $A\Rightarrow B$, there exist $a,b \in \Omega$ such that $[a]=A, [b]=B$ and $a\rightarrow b$.
We also have $a\rightarrow\ \mbox{triv}(a)$.
If we take $C=A$ and $c=\mbox{triv}(a)$ in proposition \ref{abc},
we obtain $\mbox{triv}^{-1}(b)\rightarrow \mbox{triv}(a)$, which proves that $B\Rightarrow A$.
\qed

Therefore the graph structure of the time evolution of the periodic discrete Toda equation \eqref{toda} over $\Omega$ is always {\em bidirectional}.
Thus we can write $A\Leftrightarrow B$ instead of $A \Leftarrow B$ or $A \Rightarrow B$.

Proposition \ref{abc} and (therefore) proposition \ref{bidirection} are proved to hold for a wide class of discrete equations under more general settings.
This fact will be elaborated in the last section as proposition \ref{abcgeneral} and its corollary.

\begin{Corollary}\label{poly}
Let $m\ge 2$ and $A,B_i\in\Omega\ \ (i=1,2,\cdots, m)$ be  equivalence classes distinct from each other.  
Let us assume that $m$ points $b_i\in B_i$ are all next steps of $a\in A$: i.e.,
$a\rightarrow b_i$ for all $i=1,\cdots m$. Then $B_j\Leftrightarrow B_k$ for $j\neq k$, and the points $A, B_1, \cdots, B_m$ form the complete graph $K_{m+1}$.
\end{Corollary}
$K_{m+1}$ is a graph of $(m+1)$-sided polygon with all the diagonal lines drawn.

\begin{Proposition} \label{prop4aftercor2}
For $A, B\in \Omega$, let us assume that $a\rightarrow b$ for some $a\in A$ and $b\in B$. Then (triv)$^2 a\rightarrow$(triv)$^2 b$.
\end{Proposition}
\textbf{proof}

The map $($triv$)^2$ maps $(I_1,V_1,\cdots, I_N,V_N)$
to $(I_2,V_2,\cdots, I_N,V_N,I_1,V_1)$.
Thus $($triv$)^2$ is expressed as a translation
\begin{equation}
\left\{
\begin{array}{cl}
I'_n&=I_{n+1},\vspace{2mm} \\
V'_n&=V_{n+1},
\end{array}
\right. (n=1,2,\cdots, N)
\end{equation}
which preserves Eq. \eqref{toda}.
\qed

From this proposition we learn that, to obtain all the elements of $\Omega$ which are connected to $A\in\Omega$, we need only to investigate the next steps of two points $a\in A$ and triv$(a)\in A$.
Here we can choose $a\in A$ arbitrarily.
For general $p$ and $N$, a part of the graph should be composed of several arrays of complete graphs $K_m$.
Moreover, if the initial conditions are limited to non-zero ones, then we have $m\le p-1$, since the number of next steps from a particular element $a\in A$ (where $A$ is an element of $\Omega$)
is limited to $p-1$ from proposition \ref{p-1}.
However the whole structure is not completely determined, since the number of points in $\Omega(p,N)$ increases exponentially.
We give several examples for small $p$ and $N$ in the next section.

\section{Examples}
We define the {\em order}  $\mbox{ord}\, (\mathbb{IV})$ of the state $\mathbb{IV}=(I_1,V_1,\cdots,I_N,V_N)$ as
the smallest positive integer $m>0$ such that
\[
(\mbox{triv})^m (\mathbb{IV})=(\mathbb{IV})
\]
holds. Note that $m$ should be a factor of $2N$.
Let us define $a_m$ as the number of elements with the order $m$:
\[
a_m=\#\left\{\mathbb{IV}\in(\mathbb{F}_p)^{2N}|\mbox{ord}\,(\mathbb{IV})=m \right\}.
\]
We have
\begin{equation}
\sum_{m'>0,\ m'|m} a_{m'}=p^m \label{ampm}
\end{equation}
 for every positive integer $m$ with $m|2N$, where $q|r$ indicates that $r\in\mathbb{Z}$ can be divisible by $q\in\mathbb{Z}$ in $\mathbb{Z}$.
From the recurrence relation \eqref{ampm},  we can inductively compute $a_m$ for every factor $m$ of $2N$ starting from $a_1=p$.
For example, when $N=2$, we have $a_1+a_2=p^2$ and $a_1+a_2+a_4=p^4$ from \eqref{ampm}.
Let us introduce the M\"{o}bius function defined over natural numbers.
When a positive integer $n$ is factorized as $n=p_1^{e_1}p_2^{e_2}\cdots p_k^{e_k}$, where each $p_i$ is a prime number and $e_i$ is a positive integer, then $\mu(n)$ is defined as
\[
\mu(n)=
\begin{cases}
1 & (n=1)\\
(-1)^k & (e_1=e_2=\cdots =e_k=1)\\
0 & (\mbox{otherwise})
\end{cases}.
\]
By the M\"{o}bius inversion formula, we have an explicit expression for $a_m$ for every $m|2N$:
\[
a_m=\sum_{m'>0,\, m'|m} \mu\left( \frac{m}{m'} \right) p^{m'}.
\]
From the definition of equivalence classes over $(\mathbb{F}_p)^{2N}$, we have
\[
\#\Omega(p,N)=\sum_{m>0,\ m|2N}\left(\frac{a_m}{m}\right).
\]
For example, when $N=2$, we have $\#\Omega(p,2)=a_1+a_2/2+a_4/4$.
The above discussion is valid for all initial conditions of the periodic discrete Toda equation, however, for brevity, we sometimes limit ourselves to the non-zero initial conditions in the examples below and use the following notations:
\[
b_m=\#\left\{\mathbb{IV}\in(\mathbb{F}_p^{\times})^{2N}|\mbox{ord}\,(\mathbb{IV})=m \right\},
\]
\[
\Omega' (p,N)=(\mathbb{F}_p^{\times})^{2N}/\sim,\;\; \#\Omega'(p,N)=\sum_{m>0,\, m|2N}\left(\frac{b_m}{m}\right).
\]
From the M\"{o}bius inversion formula we have
\[
b_m=\sum_{m'>0,\, m'|m}\mu\left(\frac{m}{m'}\right)(p-1)^{m'}.
\]
\subsection{$p=3$, $N=2$} \label{ex32}
For $p=3$ and $N=2$, let us first limit ourselves to non-zero initial values and classify all the sixteen elements $(I_1,V_1,I_2,V_2)\in(\mathbb{F}_p^{\times})^{2N}=\{1,2\}^4$.
We have $b_1=2,\ b_2=2,\ b_4=12$.
Thus
\[
\# \Omega' (3,2)=2+2/2+12/4=6.
\]
\[
 \Omega' (3,2)=\left\{\{40\},\ \{80\},\ \{50,70\},\ \{41,43,49,67\},\ \{53,71,77,79\},\ \{44,52,68,76\}\right\}.
\]
Here, the points in $\Omega'(3,2)$ are expressed as $10$-decimals according to the map $\phi$, 
where $\phi^{-1}(40)=(1,1,1,1)$, $\phi^{-1}(41)=(1,1,1,2)$, $\phi^{-1}(43)=(1,1,2,1)$, $\phi^{-1}(44)=(1,1,2,2)$, $\phi^{-1}(49)=(1,2,1,1)$, $\cdots$ $\phi^{-1}(79)=(2,2,2,1)$, and, $\phi^{-1}(80)=(2,2,2,2)$.
By listing all the next step of all the sixteen elements, we obtain that
all the $6$ points in $\Omega'(3,2)$ are isolated: i.e., $A\not\Rightarrow B$ for all $A, B\in\Omega'(3,2)$.
For example, all the next steps of $(1,2,1,2)$ are the two points $(1,2,1,2)$ and $(2,1,2,1)$, and all the next steps of $(2,1,2,1)$ are also $(1,2,1,2)$ and $(2,1,2,1)$.
Since triv$\,(1,2,1,2)=(2,1,2,1)$, these two points are in the same class in $\Omega'(3,2)$.
($\phi(1,2,1,2)=50$, $\phi(2,1,2,1)=70$.)
Therefore $[(1,2,1,2)]\in\Omega'(3,2)$ is isolated.

Next, we study the whole space $(\mathbb{F}_3)^4$. We have $a_1=3$, $a_2=6$, $a_4=72$ and $\#\Omega(3,2)=3+6/2+72/4=24$.
By investigating all the transitions of $81$ points $(I_1,V_1,I_2,V_2)$ in $(\mathbb{F}_3)^4$ by the discrete Toda Eq. \eqref{toda},
we conclude that $\Omega(3,2)$ consists of the following $24$ points:
\begin{itemize}
\item $1$ set of $P_3$ $(3$ points$)$,
\item $5$ sets of $K_2$'s $(10$ points$)$,
\item $11$ isolated points.
\end{itemize}
Here, the symbol $K_m$ $(m\ge 3)$ denotes the complete graph and we have used the notation $K_2$ denote a line segment. $P_3$ denotes a triangular graph of three $K_2$'s.
We will denote the decomposition of the graph $\Omega(3,2)$ into connected components as
\[
\Omega(3,2)=P_3+5K_2+K_1,
\]
where $K_1$ denotes an isolated point. For a graph $G$ and an integer $k$, the notation $kG$ denotes that
we have $k$ connected components $G$.
Note that $K_3$ denotes the same triangular graph as $P_3$, since a triangle does not possess a diagonal line.

\begin{figure}
\centering
\includegraphics[width=12cm,bb=70 250 500 800]{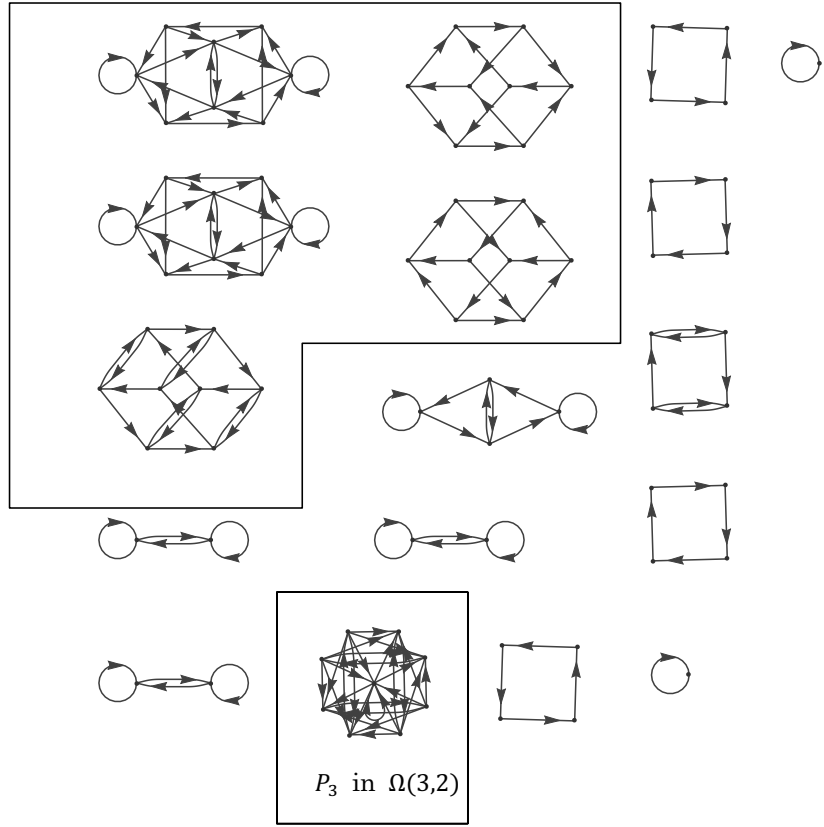}
\caption{Graph of solutions to the discrete Toda Eq. \eqref{toda} for $p=3$ and $N=2$. All the possible evolutions to and from the $81$ points in $(\mathbb{F}_3)^4$
are shown. Coordinates at the vertices are omitted. Correspondence with the equivalence classes in $\Omega(3,2)$ is shown on the figure. Five graphs framed on the left corresponds to $K_2$ (line segment), and the graph at the bottom to $P_3$ (triangle).
Other connected components correspond to isolated points in $\Omega(3,2)$.}
\label{fig5}
\end{figure}

As a comparison, we give a graph of solutions to Eq. \eqref{toda} over $(\mathbb{F}_3)^4$ in Fig. \ref{fig5}.
All the possible evolutions to and from the $81$ points in $(\mathbb{F}_3)^4$ are shown.
In the figure, a square denotes a shift transformation (which we call as a trivial transition)
\[
(I_1,V_1,I_2,V_2)\to (V_1,I_2,V_2,I_1)\to (I_2,V_2,I_1,V_1)\to (V_2,I_1,V_1,I_2)\to (I_1,V_1,I_2,V_2).
\]
If the state $(I_1,V_1,I_2,V_2)$ has an additional symmetry (e.g. $(1,2,1,2)$, $(1,1,1,1)$), the square collapses into a line or a point.
We can understand that taking the equivalence class by the trivial transition $($triv$)$ is important in observing the essential structures
of the solution to Eq. \eqref{toda}.

\subsection{$p=5$, $N=3$}
\begin{figure}
\centering
\includegraphics[width=8cm,bb=75 500 300 750]{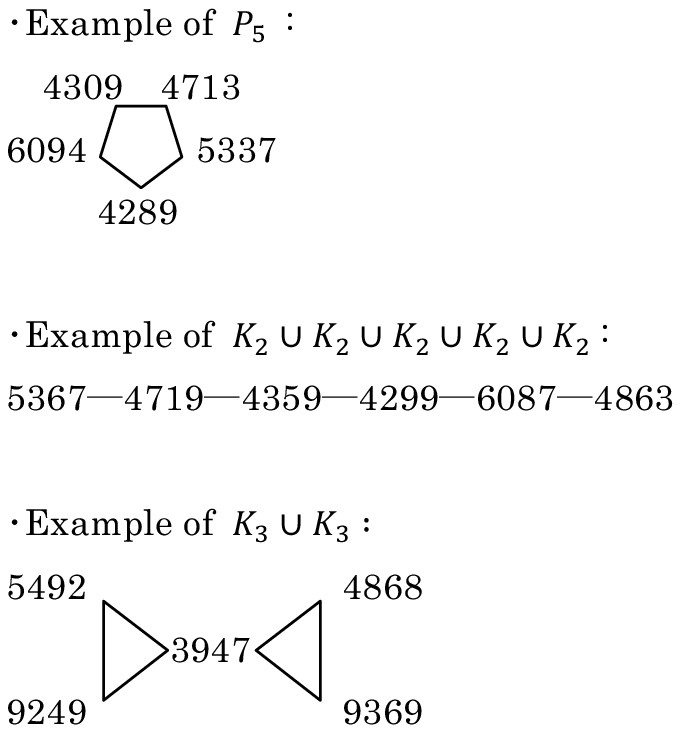}
\caption{Examples of connections of elements of $\Omega'$ when $p=5,N=3$.
An edge denotes a connection between two equivalence classes in $\Omega'(5,3)$: e.g.,  $[4309]\Leftrightarrow [4713]$.
We have picked up the structures $P_5$, $K_2\cup K_2\cup K_2\cup K_2\cup K_2$
 and $K_3\cup K_3$. The numbers at the vertices indicate the representative elements of the equivalence class in $\Omega'(5,3)\subset S/ \sim$.}
\label{fig2}
\end{figure}

For $p=5$ and $N=3$, let us classify all the $4^6=4096$ elements $(I_1,V_1,I_2,V_2,I_3,V_3)\in(\mathbb{F}_p^{\times})^{2N}=\{1,2,3,4\}^6$.
We have $b_1=4,\ b_2=12,\ b_3=60$ and $b_6=4020$.
Thus $\#\Omega' (5,3)=4+12/2+60/3+4020/6=700$.
Numerical calculations show that these $700$ points are classified as follows:
\begin{itemize}
\item $4$ sets of $K_3\cup K_3$'s $(20$ points$)$.
\item $4$ sets of $P_5$'s $(20$ points$)$.
\item $14$ sets of $K_3$'s $(42$ points$)$.
\item $2$ sets of $K_2\cup K_2\cup K_2\cup K_2\cup K_2$'s $(12$ points$)$.
\item $4$ sets of $K_2\cup K_2\cup K_2\cup K_2$'s $(20$ points$)$.
\item $12$ sets of $K_2\cup K_2\cup K_2$'s $(48$ points$)$. 
\item $26$ sets of $K_2\cup K_2$'s $(78$ points$)$.
\item $82$ sets of $K_2$'s $(164$ points$)$.
\item The remaining $296$ isolated points.
\end{itemize}
A symbol $P_5$ denotes a pentagonal graph composed of five $K_2$'s. Thus we obtain
\[
\Omega'(5,3)=4(K_3\cup K_3)+4P_5+14K_3+2(K_2)^5+4(K_2)^4+12(K_2)^3+26(K_2)^2+82K_2+296K_1.
\]
Here, for a line segment $K_2$ and a positive integer $k$,
the graph $(K_2)^k$ denotes a polyline $\underbrace{K_2\cup K_2 \cup\cdots \cup K_2}_{k}$.
We give examples of the connections between the equivalence classes (elements of $\Omega' (5,3)\subset S/\sim$). 
For a class $A\in S / \sim$, we take a representative element $\alpha\in S$ so that $[\alpha]=A$ and locate $\alpha$ at the vertex of the graph.
If $\#A>1$, the element $\alpha\in S$ is chosen so that $\alpha$ is the smallest element in $A$.
The graphs having three or more sides from one vertex arise from the special solutions which appear only over finite fields, as we have seen in the first case of the proposition \ref{p-1}.
For example, the point $3947\in S$ in the graph $K_3\cup K_3$ in Fig. \ref{fig2} has four sides. In fact
\[
\phi^{-1}(3947)=(1,1,1,2,4,2)\in\left(\mathbb{F}_5^{\times}\right)^6,
\]
and for $(I_1^t,V_1^t,I_2^t,V_2^t,I_3^t,V_3^t)=(1,1,1,2,4,2)$, we have $\alpha^t=1$ and $X_1^t = X_2^t = X_3^t= 0$ $\mod 5$.

\subsection{Larger $p$ and $N$}
\begin{figure}
\centering
\includegraphics[width=12cm,bb=75 550 530 750]{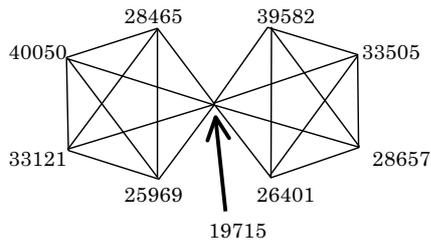}
\caption{An example of connections of elements of $\Omega$ when $p=7,N=3$. We see the structure
$K_5\cup K_5$.}
\label{fig3}
\end{figure}

\begin{figure}
\centering
\includegraphics[width=12cm,bb=75 480 530 750]{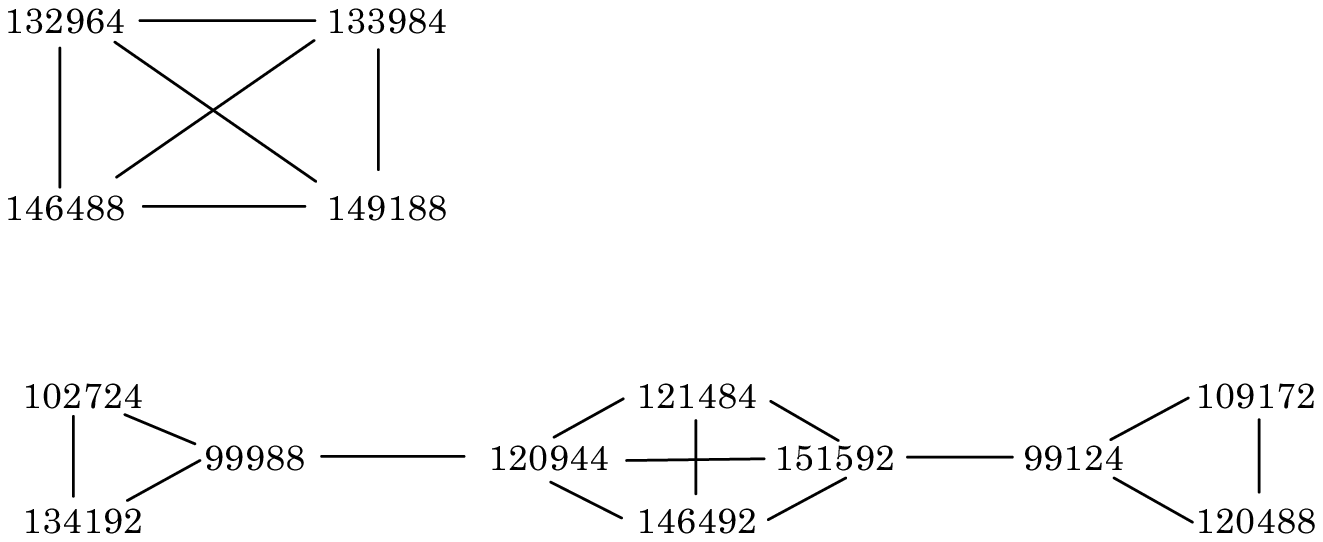}
\caption{Examples of connections of elements of $\Omega$ when $p=5,N=4$. We see the structures
$K_4$ and $K_3\cup K_2\cup K_4\cup K_2\cup K_3$.}
\label{fig4}
\end{figure}

We present more complex examples of graphs of equivalence classes for $p=7,N=3$ and $p=5,N=4$ in Fig. \ref{fig3} and Fig. \ref{fig4}. Investigation all the classes for these parameter values is practically impossible because of the number of possible initial states increases exponentially ($7^6=117649$, $5^8=390625$).

\subsection{General $p$ and $N=2$}
For $N=2$ we will prove that $\Omega(p,2)$ is composed only of the following items: one $K_p$, several $K_m$'s $(4\le m\le p-1)$, $P_m$'s $(3\le m\le p-1)$, $(K_2)^m:=\underbrace{K_2\cup K_2\cup \cdots \cup K_2}_{m}$'s $(2\le m\le p-2)$, and isolated points. In other words, a graph with more than one complete graph $K_m$ $(m\ge 3)$ as a subgraph does not appear. More precisely, we have
\begin{Theorem}\label{conj2}
The decomposition into the connected components of the graph of $\Omega(p,2)$ and $\Omega'(p,2)$ are given as follows:
for $p=2, 3$,
\begin{eqnarray*}
\Omega(2,2)&=&K_2+4K_1,\; \Omega'(2,2)=K_1\\
\Omega(3,2)&=&K_3+5K_2+11K_1,\; \Omega'(3,2)=6K_1,
\end{eqnarray*}
and for $p\ge 5$, we have
\[
\Omega(p,2)=K_p+(p-1)K_{p-1}+(p-1)K_{(p-1)/2}+\sum_{k=3}^{n} b_k P_{k}+\sum_{k=1}^{m} c_k (K_2)^k+ d K_1,
\]
\[
\Omega' (p,2)=\sum_{k=3}^{n} b'_k P_{k}+\sum_{k=1}^{m} c'_k (K_2)^k+ d' K_1,
\]
where $n, m$ are some positive integers which depend only on $p$, and the coefficients $b_k\ge b'_k$, $c_k\ge c'_k$, $d\ge d'$ are non-negative integers.
Here, for graphs $A,B,C$, the sum $A=B+C$ denotes that $A$ has two connected components $B$ and $C$.\end{Theorem}
\textbf{proof}\;\;
In the case of $p=2,3$ the assertion is already proved (we refer to example \ref{ex32} for $p=3$).
To prove the latter half of theorem \ref{conj2}, we prepare the following lemma \ref{n2lemma} on the number of next steps:
\begin{Lemma}\label{n2lemma}
The next step of $(a,b,c,d)\in(\mathbb{F}_p)^4$ with respect to Eq.~\eqref{toda} is classified as follows:
\end{Lemma}
\begin{enumerate}
\item If $a+b\neq 0$ and $c+d\neq 0$ and $ac \neq bd$ then the next steps of $(a,b,c,d)$ are the two states
$(b,c,d,a)$ and
\[
\left(\frac{c(a+b)}{c+d}, \frac{b(c+d)}{a+b}, \frac{a(c+d)}{a+b}, \frac{d(a+b)}{c+d}\right).
\]
\item If $a+b\neq 0$ and $c+d\neq 0$ and $ac=bd$ then $(a,b,c,d)$ has only one next step $(b,c,d,a)$.
\item If $a+b\neq 0$ and $c+d=0$ and $d\neq 0$ then $(a,b,c,d)$ has only one next step $(b,c,d,a)$.
\item If $a+b\neq 0$ and $c=d=0$  then $(a,b,c,d)$ has $p$ next steps: \label{n2lemc4}
\[
(a+b-x,0,0,x),\;\;x=0,1,\cdots,p-1.
\]
\item If $a+b=0$ and $c+d\neq 0$ and $a\neq 0$ then $(a,b,c,d)$ has only one next step $(b,c,d,a)$.
\item If $a=b=0$ and $c+d\neq 0$ then $(a,b,c,d)$ has $p$ next steps: \label{n2lemc6}
\[
(0,x,c+d-x,0),\;\;x=0,1,\cdots,p-1.
\]

\item If $a+b=0$ and $c+d= 0$ and $ac\neq 0$ then $(a,b,c,d)$ has $p-1$ next steps: \label{n2lemc7}
\[
\left(x,-\frac{ac}{x},\frac{ac}{x},-x\right),\;\; x=1,2,\cdots ,p-1.
\]
\item  If $a+b=0$ and $c+d= 0$ and $ac= 0$ then $(a,b,c,d)$ has $2p-1$ next steps: \label{n2lemc8}
\[
(x,0,0,-x),\ (0,-x,x,0),\ (0,0,0,0),\;\;\;x=1,2,\cdots ,p-1.
\]
\end{enumerate}
For an evolution $(a,b,c,d) \rightarrow (a',b',c',d')$ of Eq.~\eqref{toda}, we obtain
\begin{align*}
&\{(a+b)c'-(c+d)a\}(c'-a)=0,\\
&\{(c+d)a'-(a+b)c\}(a'-b)=0.
\end{align*}
Using these relations, we can obtain lemma \ref{n2lemma} by elementary computation.

We now utilize lemma \ref{n2lemma} to obtain the graph structure of $\Omega(p,2)$.
First let us consider the case \ref{n2lemc8}. We have an evolution of Eq.~\eqref{toda} from $(0,0,0,0)$ to all the $2p-1$ elements $(x,0,0,-x),(0,-x,x,0),(0,0,0,0)\in(\mathbb{F}_p)^4$.
Since $(x,0,0,-x)=($triv$)^2(0,-x,x,0)$ for every $x\in\mathbb{F}_p$, these $2p-1$ elements form $p$ classes $[(x,0,0,-x)]$ $(x=0,1,2,\cdots,p-1)$ in $\Omega(p,2)$.
Therefore the case \ref{n2lemc8} constitutes one complete graph $K_p$ in $\Omega(p,2)$ from corollary \ref{poly}.
Next let us study the case \ref{n2lemc6}. The $p$ next steps $(0,x,c+d-x,0)$ of the state $(0,0,c,d)$ $(c+d\neq 0)$ consists of $p-1$ classes in $\Omega(p,2)$, since $(0,c+d,0,0)\sim (0,0,c+d,0)$ belong to the same class.
We note that $(0,0,c,d)\sim (0,x,c+d-x,0)$ when $x=c$. Thus the graph which includes the vertex $[(0,0,c,d)]$ constitutes  the complete graph $K_{p-1}$ in $\Omega(p,2)$ again from corollary \ref{poly}.
We have exactly $p-1$ number of $K_{p-1}$'s, depending on the values of $c+d\in\mathbb{F}_p^{\times}$.
The case \ref{n2lemc4} forms the same graphs as in the case \ref{n2lemc6}, since $(x,y,0,0)\sim (y,0,0,x)$ for every $x,y\in\mathbb{F}_p$.
The next is the case \ref{n2lemc7}. The set $\{(x,-ac/x,ac/x,-x)\, |\, x=1,2,\cdots,p-1\}$ $(ac\neq 0)$ constitutes $(p-1)/2$ elements in the set of equivalence classes $\Omega(p,2)$ from the symmetry with respect to $($triv$)$.
Since these $(p-1)/2$ classes include the original one $[(a,-a,c,-c)]$, they form the complete graph $K_{(p-1)/2}$.
Two classes $[(a,-a,c,-c)]$ and $[(a',-a',c',-c')]$ are in the same graph $K_{(p-1)/2}$ if and only if $ac=a'c'$. Thus we have $(p-1)$ number of $K_{(p-1)/2}$'s.
Other cases do not produce a complete graph $K_m$ with $m\ge 4$. Since the number of next steps is limited to at most $2$ in $(\mathbb{F}_p)^4$, there exist at most one side from each point $(a,b,c,d)\in(\mathbb{F}_p)^4$ aside from a trivial transition obtained by (triv). Thus there should be at most two sides from a class $A$ in $\Omega(p,2)$, one is from a point $a\in A$ and the other is from triv$(a)\in A$. Note that we can take $a$ in $A$ arbitrarily from the comments after proposition \ref{prop4aftercor2}. Therefore we have only polylines $(K_2)^k$ for some integer $k\ge 1$.
\qed
For example, when $p=5$, we have
\begin{eqnarray*}
\Omega(5,2)&=& K_5+4K_4+4P_3+4(K_2)^3+8(K_2)^2+25K_2+42K_1,\\
\Omega'(5,2)&=& 4(K_2)^2+15K_2+28K_1.
\end{eqnarray*}
When $p\ge 7$, we have
\begin{eqnarray*}
\Omega(7,2)&=& K_7+6K_6+6P_4+24P_3+12(K_2)^5+6(K_2)^4+18(K_2)^3+24(K_2)^2+60 K_2+111 K_1,\\
\Omega'(7,2)&=&
18P_3+12(K_2)^3+24(K_2)^2+39K_2+84K_1,\\
\Omega(11,2)&=& K_{11}+10K_{10}+10K_5+10P_6+20P_5+\cdots+20(K_2)^9+\cdots,\\
\Omega(13,2)&=& K_{13}+12K_{12}+12K_6+36P_7+12P_6+60P_5+\cdots+36(K_2)^{11}+\cdots.
\end{eqnarray*}
From these observations, we conjecture that we can slightly refine proposition \ref{conj2}:
\begin{Conjecture}
The decomposition into the connected components of the graph of $\Omega(p,2)$ and $\Omega'(p,2)$ for $p\ge 5$ are given as
\[
\Omega(p,2)=K_p+(p-1)K_{p-1}+(p-1)K_{(p-1)/2}+\sum_{k=3}^{(p+1)/2} b_k P_{k}+\sum_{k=1}^{p-2} c_k (K_2)^k+ d K_1,
\]
\[
\Omega' (p,2)=\sum_{k=3}^{(p+1)/2} b'_k P_{k}+\sum_{k=1}^{p-2} c'_k (K_2)^k+ d' K_1,
\]
where the coefficients $b_k\ge b'_k$, $c_k\ge c'_k$, $d\ge d'$ are non-negative integers.
\end{Conjecture}
We conjecture that only $P_n$ with $n\le (p+1)/2$ and $(K_2)^m$ with $m\le p-2$ appear as connected components of $\Omega(p,2)$.

\section{General systems}
Lastly, we discuss when the network graphs of possible evolutions can be well-defined over the space of equivalence classes and is bi-directional.
We study a general discrete system of $N$ variables $u_1^t, u_2^t, \cdots ,u_N^t$, which is defined by the following simultaneous equations
\begin{equation}
f_i(u_1^t, u_2^t, \cdots ,u_N^t)=g_i(u_1^{t+1}, u_2^{t+1}, \cdots ,u_N^{t+1}),\ \ (i=1,2,\cdots, k), \label{general}
\end{equation}
where $k\ge 1$ and $f_i, g_i$ are polynomials.
We impose the periodic boundary condition $u_i^t=u_{i+N}^t$ $(\forall i\,  \forall t)$.
For example, the discrete Toda equation \eqref{toda} is obtained if we take $k=2N$, $N\to 2N$, and $f_{n}(u_1,\cdots, u_{2N})=u_{2n-1}+u_{2n}$,
$g_{n}(u_1,\cdots ,u_{2N})=u_{2n-2}+u_{2n-1}$, $f_{n+N}(u_1,\cdots, u_{2N})=u_{2n} u_{2n+1}$, $g_{n+N}(u_1,\cdots, u_{2N})=u_{2n-1}u_{2n}$, where $n=1,2,\cdots, N$, and the subscripts of $u$ are counted modulo $2N$.
We use the same notations as in the previous sections for this system: e.g.,
we write $(u_1^t, u_2^t, \cdots ,u_N^t)\to (u_1^{t+1}, u_2^{t+1}, \cdots ,u_N^{t+1})$ when both \eqref{general} and the boundary condition are satisfied, and we use the mapping (triv)$(u_1^{t}, u_2^{t}, \cdots ,u_N^{t})=(u_2^{t}, u_3^{t},  \cdots ,u_N^{t}, u_1^t)$, and the set of equivalence classes $\Omega=(\mathbb{F}_p)^N/\sim$.
We prove that the same statement as that of proposition \ref{abc} holds if (triv) is one of the solutions of the system \eqref{general}.
\begin{Proposition} \label{abcgeneral}
Suppose that
\begin{equation}
f_i(u_1^{t}, u_2^{t}, \cdots ,u_N^{t})=g_i\left((triv)(u_1^{t}, u_2^{t}, \cdots ,u_N^{t})\right), \label{trivcond}
\end{equation}
for all $(u_1^t,u_2^t,\cdots, u_N^t)\in(\mathbb{F}_p)^N$.
Then, for any $A,B,C\in\Omega$, if there exist $a\in A$, $b\in B$ and $c\in C$ such that $a\to b$ and $a\to c$ hold, then we have (triv)$^{-1}(c)\to b$.
\end{Proposition}
\textbf{Proof}\;\;
First we have $f_i\circ (\text{triv})=g_i$ for every $i=1,2,\cdots, k$, from the condition \eqref{trivcond}.
Since $a\to b$, we have $f_i(a)=g_i(b)=\left(f_i\circ (\text{triv})^{-1}\right)(b)$ for all $i$. In the same manner we have $f_i(a)=\left(f_i\circ (\text{triv})^{-1}\right)(c)$ for all $i$ from $a\to c$.
Therefore
\[
f_i\left((\text{triv})^{-1} (b)\right)=\left(f_i\circ (\text{triv})^{-1}\right) (c)=g_i(c),
\]
which proves that $(\text{triv})^{-1}(b)\to c$.
\qed

\begin{Corollary}
For every $A,B\in\Omega$, if $A\Rightarrow B$ then $B\Rightarrow A$.
\end{Corollary}
We have obtained that the graph structure over $\Omega$ is bi-directional under the generalization in this section.

Let us note that we can further ease the condition \eqref{trivcond} by replacing the map (triv) with an arbitrary permutation $\sigma\in\mathfrak{S}_N$ and by supposing that $f_i\circ \sigma =g_i$ for all $i=1,2,\cdots, k$.
We can prove in the same manner that the graphs over $\tilde{\Omega}:=(\mathbb{F}_p)^N/ \sim' $ is bi-directional, where $\sim'$ is an equivalence class induced from the action of $\sigma$ on $(\mathbb{F}_p)^N$.
We conclude that the property of bi-directionality of the graphs is satisfied for a wide class of discrete mappings.
Let us also note that the results in this section are not limited to equations over finite fields, and are applicable to equations over any field.

\section{Conclusion}
In this paper we have studied the periodic discrete Toda equation, and established the structures of all the possible solutions over finite fields.
Since we have multiple choice of solutions, we cannot uniquely determine the time evolution in an usual sense.
Instead, we have constructed the finite graph structures over the space of states, by drawing arrows whenever the equation is satisfied.
Moreover, we have shown that, if we introduce the equivalence classes by identifying the states which can be reached by using cyclic permutations (trivial evolutions),
the quotient space of the states constitute bidirectional graphs.
Therefore we do not have to consider the direction of the arrows.
We have proved that the graphs satisfy a weak type of transitive relation, and therefore consist of several arrays of complete graphs and line segments. We have discussed the number of possible sides allowed for polygons: the upper limit of the sides is $p-1$ for non-zero states, but not limited to $p-1$ for the states with at least one $0$.
One of the future problems is to completely determine the structure of the graphs for general $p$ and $N$, and for the states with $I_n^t=0$ or $V_m^s=0$ for some $n,m,t,s$.
Since the graphs over the simplified space of initial conditions can be bi-directional in a wide class of discrete systems, application of our methods to other discrete integrable equations and cellular automata should be possible without major obstacles. We hope that the study of discrete integrable equations over finite fields will provide effective tools for analyzing mathematical models of various phenomena in engineering sciences.

\section*{Acknowledgments}
Authors wish to thank Prof. Ralph Willox for useful comments.
This work was partially supported by JSPS KAKENHI Grant Number 15H06128.


\end{document}